\documentclass{article}
\usepackage[utf8]{inputenc}
\usepackage{amsmath,amssymb}
\usepackage{appendix}
\usepackage{slashed}
\usepackage{comment}
\usepackage{hyperref}
\usepackage{ytableau}
\usepackage{dsfont}
\usepackage{cite}
\usepackage{tensor}
\usepackage{multicol}
\usepackage{authblk}
\usepackage{verbatim}

\evensidemargin=\oddsidemargin
\title{The chiral torsional anomaly and the Nieh-Yan invariant with and without boundaries}
\author[a]{Johanna Erdmenger\thanks{ johanna.erdmenger@uni-wuerzburg.de}}
\author[b]{Ioannis Matthaiakakis\thanks{I.Matthaiakakis@soton.ac.uk}}
\author[a]{Ren\'{e} Meyer\thanks{ rene.meyer@uni-wuerzburg.de}}
\author[c]{Dmitri Vassilevich\thanks{dvassil@gmail.com}}

\affil[a]{Institute for Theoretical Physics and Astrophysics, and W{\"u}rzburg-Dresden Cluster of Excellence on Complexity and Topology in Quantum Matter ct.qmat, Julius-Maximilians-Universit{\"a}t W{\"u}rzburg, Am~Hubland, D-97074~W{\"u}rzburg, Germany}
\affil[b]{Mathematical Sciences and STAG Research Centre, University of Southampton, Highfield,
Southampton SO17 1BJ, United Kingdom}
\affil[c]{Centro de Matem{\'a}tica, Computa\c{c}\~{a}o e Cogni\c{c}\~{a}o - Universidade Federal do ABC, Santo Andr\'{e}, SP, Brazil}
\date{\today}

\usepackage{xcolor}

\newcommand{\MM}{\mathcal{M}}
\newcommand{\nn}{\mathbf{n}}
\newcommand{\dd}{\mathrm{d}}

\begin{document}

\maketitle
\begin{abstract}
    There exists a long-standing debate regarding the torsion contribution to the 4d chiral anomaly of a Dirac fermion. Central to this debate is the Nieh-Yan anomaly, which has been considered ill-defined and a regularization artifact. Using a heat-kernel approach, we examine the relationship between the Dirac operator index, the Nieh-Yan invariant and the torsional anomaly. We show the Nieh-Yan invariant vanishes on spacetimes without boundaries, if the Dirac index is well-defined. In the known examples of non-vanishing Nieh--Yan invariant on manifolds without boundaries, the heat kernel expansion breaks down, making the index ill-defined. Finally, for finite boundaries we identify several finite bulk and boundary anomaly terms, alongside bulk and boundary Nieh-Yan terms. We construct explicit counterterms that cancel the Nieh-Yan terms and argue that the boundary terms give rise to a torsional anomalous Hall effect. Our results emphasize the importance of renormalization conditions, as these can affect the non-thermal Nieh-Yan anomaly coefficients. In addition, we demonstrate  that anomalous torsional transport may arise even without relying on the Nieh-Yan invariant.
\end{abstract}

\section{Introduction}

In quantum field theory, anomalies signal the breakdown of a classical charge conservation equation. They have played a major role in the development of high-energy physics and the standard model \cite{Peskin:1995ev, Bertlmann:1996xk, Kiritsis:2019npv,Bonora:2023soh}, and  have also emerged as an essential feature of low-energy condensed matter, strongly influencing its transport properties \cite{Landsteiner:2016led, Chernodub:2021nff}. In the present paper we re-examine a debate, present in both the high energy and condensed matter literature, concerning the chiral $U(1)$ anomaly of a single $4d$ Dirac fermion generated by torsion, henceforth referred to as the chiral torsional anomaly. 

Torsion is a property of spacetime on a similar footing to curvature, in the sense that it can be measured by parallel transporting vectors along closed spacetime loops. In high-energy settings, torsion is generated by the spin of quantum matter, although it is expected to be negligible away from early universe physics \cite{RevModPhys.48.393,Shapiro:2001rz}. The situation is different in condensed matter, where torsion can be generated in principle at will, by appropriate lattice deformations, and appears naturally in low-energy descriptions of tight-binding models \cite{doi:10.1142/0356,KATANAEV19921,NelsonDefects}.

The debate regarding the chiral torsional anomaly began nearly three decades ago in the high-energy domain, when calculations of the chiral torsional anomaly gave rise to two distinct and seemingly incompatible results. Namely, the discussion \cite{Chandia:1997hu,Obukhov:1997pz,Kreimer:1999yp} (see also \cite{Nascimento_2022}) has been focused on the presence or not of a single anomaly term, the one proportional to the Nieh-Yan (NY) invariant \cite{Nieh:1981ww}
\begin{equation}
\label{Eq:NYInv}
C_{\rm NY} = {1\over 4}\epsilon^{\mu\nu\rho\lambda}\left(\eta_{\alpha\beta}T^\alpha_{[\mu\nu}T^\beta_{\rho\lambda]}-2R_{\mu\nu\lambda\rho} \right) = \nabla_\mu \left({1\over 3!}\epsilon^{\mu\nu\rho\lambda}T_{[\nu\rho\lambda]}\right)~,
\end{equation}
with $T^a_{\mu\nu}, R_{\mu\nu\rho\lambda}$ the torsion and curvature tensors, and $\epsilon^{\mu\nu\rho\lambda}, \eta_{ab}$ the Levi-Civita tensor and Minkowski metric, respectively. Antisymmetrization is defined with a unit factor, e.g. $T_{[\mu\nu]} = T_{\mu\nu}- T_{\nu\mu}$ and similarly for all higher order tensors.

The debate has focused on two features of the NY contribution to the anomaly, which we refer to as the NY term for brevity. First, the NY term appears in the chiral charge conservation equation with a coefficient, which plays the role of the cutoff in a cutoff regularization scheme. This goes against the established lore of quantum anomalies, whose coefficients are quantized or severely constrained \cite{Bertlmann:1996xk}. Anomalies involving background fields generically appear with dimensionless coefficient in renormalizable quantum field theories. The presence of the cutoff also contradicts calculations in different regularization schemes, where the NY term is absent \cite{Yajima:1985jd,Kimura:2007xa,Mavromatos:1987ru}. 

The second feature of the NY term which has been a subject of debate is the NY invariant itself. We  see from \eqref{Eq:NYInv} that $C_{\rm NY}$ is a total divergence. This is true for all contributions to the chiral torsional anomaly, stemming also from other background fields, with one crucial difference: The NY invariant can be expressed as a divergence globally, not just in a restricted patch of spacetime. As a result, upon integrating $C_{\rm NY}$ over Minkowski spacetime with fall-off conditions on the torsion, we find a vanishing result for the global NY invariant. As a counterargument to this, explicit examples of spacetimes with a non-trivial $C_{\rm NY}$ integral have been constructed, following techniques familiar from the magnetic monopole literature \cite{Obukhov:1997pz}. 

On the condensed matter side, the debate has focused mainly on the nature of the cutoff scale. There the cutoff dependence is seen as less problematic, since a physical cutoff is always a part of a condensed matter system definition. This has led to a plethora of papers predicting the NY term in semimetals and its consequences on charge transport, as well as devising methods to measure them in experiment  \cite{Sumiyoshi:2015eda,Khaidukov:2018oat,Parrikar:2014usa,Hughes:2012vg,Ferreiros:2018udw,Nissinen:2019mkw,Nissinen:2019wmh,Huang:2019adx,Huang:2019haq,Laurila:2020yll,Imaki:2020csc}. In addition, putting Dirac fermions at finite temperature has yielded a thermal contribution to the NY term \cite{Nissinen:2019mkw,Nissinen:2019wmh,Huang:2019haq}. This contribution is deemed universal, in the sense that it does not depend on the UV structure of the theory. This way we also depart from the typical lore, which sees anomalies as purely UV effects. As in high-energy, so in condensed matter, the NY term has been argued to be a renormalization scheme artifact with no effect on low-energy transport physics \cite{Ferreiros:2020uda, Chernodub:2021nff}.

In this paper, we reconsider the heat kernel formula for calculating the index and the related chiral torsional anomaly. According to common beliefs, integrating the anomaly over the base manifold yields a topological invariant of the manifold, the index of the Dirac operator. One may thus conclude \cite{Chandia:1997hu,Chandia:1997jf,Chandia:1998nu,Rasulian:2023gtb} that in the presence of the NY term, the index includes the NY invariant and inherits the cutoff dependence\footnote{To remove this dependence, it was proposed \cite{Chandia:1997hu,Chandia:1997jf,Chandia:1998nu} to make a cutoff-dependent rescaling of the anomaly. We comment on this trick in Section \ref{sec:ind} below.}. This cannot be true. We resolve this contradiction by showing that the conditions for having a well-defined index do not hold for field configurations considered in \cite{Chandia:1997hu,Chandia:1997jf,Chandia:1998nu,Rasulian:2023gtb}. Thus the NY term does not contribute to the index of a torsionful Dirac operator in spacetimes with non-compact boundaries.

We also consider the case of compact boundaries, which can in principle host a non-trivial NY invariant. Compact boundaries are also a natural setup from the condensed matter point of view. We show that the NY term can be cancelled by finite counterterms, indicating the  coefficient of the NY term depends on the renormalization conditions imposed on the fermion interaction parameters. In addition, we find several finite terms both in the bulk and boundary with the following properties. The bulk terms consist of the usual Riemannian curvature anomaly along with a Pontryagin term for the first derivative of torsion. The anomaly on the boundary contains a Chern-Simons term for torsion, alongside with several non-topological terms. The interpretation of the non-topological terms remains unclear, but we argue that the Chern-Simons term leads to an anomalous torsional Hall effect in a domain wall geometry. We present an explicit formula for the corresponding torsional Hall conductivity in Eq.~\eqref{Eq:BoundaryCurrent}.

We begin our discussion with section \ref{sec:Geometry}, where we set up our conventions for the geometry and Dirac operator we use. In section \ref{sec:ind}, we re-examine the conditions for having a well-defined Dirac operator index and the corresponding anomaly. In section \ref{sec:chian}, we derive the chiral torsional anomaly for spacetimes with boundaries and some of their effects on transport. We conclude the paper with a discussion of our results and future directions in section \ref{sec:concl}. To make our paper self-contained we included some background material which, together with details of (rather standard) calculations, is included  appendices.

\section{Riemann-Cartan geometry and Dirac action}\label{sec:Geometry}

In this section, we specify the background geometry and corresponding Dirac action and operator, which we use to calculate the chiral torsional anomaly. 

We consider an $n$-dimensional manifold ${\cal M}$, with $n$ even, equipped with a Euclidean metric $g_{\mu\nu}$ and an independent linear connection $\Gamma^\lambda_{\mu\nu}$. The connection is assumed to be metric, with non-zero torsion and curvature. Thus ${\cal M}$ is a Riemann-Cartan spacetime \cite{RevModPhys.48.393}. To allow for the propagation of fermions on $\MM$, we assume that it admits a local inertial frame and its inverse\footnote{Index convention: Letters from the beginning of the greek alphabet, $\alpha,\beta,\dots$, denote inertial frame indices, letters from the middle of the greek alphabet, $\mu,\nu,\dots,$ denote spacetime indices, and latin letters denote boundary indices.}
$$
e^\alpha = e^\alpha_\mu dx^\mu ~,~ E_\alpha = E^\mu_\alpha \partial_\mu~,~ \alpha,\mu=1,2,\dots,n~,
$$
such that
$$
g_{\mu\nu} = \delta_{\alpha\beta}e^\alpha_{\mu} e^\beta_{\nu} ~,~e^\alpha(E_\beta) =  e^\beta_\mu E^\mu_\alpha = \delta^\alpha_\beta~.
$$ 
Apart from the metric, we can also pull the connection to the local inertial frame and define the spin connection $\omega^{\alpha}_{~\beta} = \omega^\alpha_{\mu \beta}dx^\mu$ via
\begin{equation}
\label{Eq:SpinConnections}
\partial_\mu e^\alpha_\rho + \omega^\alpha_{\mu \beta}e^\beta_\rho -e^\alpha_\lambda \Gamma^\lambda_{\mu\rho} = 0 \Leftrightarrow \omega^\alpha_{\mu \beta} = e^\alpha_\nu \partial_\mu E^\nu_\beta + e^\alpha_\nu\Gamma^\nu_{\mu\lambda}E^\lambda_\beta,
\end{equation}
The connection $\omega$ is a covector under diffeomorphisms, but transforms as a connection under local $SO(n)$ transformations. 

We consider a (Euclidean) Dirac fermion $\psi$ propagating on this spacetime. To define it, we introduce the ``inertial'' gamma matrices $\gamma^\alpha$ satisfying the Euclidean Clifford algebra
\begin{equation}
\label{gammagamma}
\lbrace \gamma^\alpha, \gamma^\beta\rbrace = 2\delta^{\alpha\beta}~,
\end{equation}
which we can push to spacetime gamma matrices via $\gamma^\mu = E^\mu_\alpha \gamma^\alpha$. In terms of the inertial gammas we can also define the $SO(n)$ generators of the Dirac spinor representation,
\begin{equation}
\label{Eq:EuclidGenerators}
\Sigma^{\alpha\beta} = {1\over 4 }[\gamma^\alpha, \gamma^\beta]~.
\end{equation}

\noindent Finally, we can couple our Dirac fermion to the background gauge fields and geometry by introducing the spinor covariant derivative 
\begin{equation}
 \label{Eq:CovDer}  
 D_\mu = \partial_\mu + \omega_\mu + i A_\mu + A_{ 5\mu}\gamma_*~.
\end{equation}
We have introduced $A_\mu$ and $A_{5\mu }$, the U(1) and U(1) chiral gauge fields, respectively, as well as $\omega_\mu = {1\over 2}\omega^{\alpha\beta}_\mu\Sigma_{\alpha\beta}$ and  $\gamma_* \equiv (-i)^{n/2}\gamma^1\gamma^2\dots\gamma^n$ the spin connection and the Euclidean $n$th gamma matrix respectively. With these conventions in place, we can write down an action for $\psi$
\begin{equation}
    \label{Eq:DiracActionEucl}
    S_E = \int_{\cal M} d^n x |e|\psi^\dagger E^\mu_{\alpha}\gamma^\alpha i D_\mu \psi \equiv~\int_{\cal M} d^n x |e|\psi^\dagger\slashed{D} \psi,
\end{equation}
with $|e|$ the determinant of $e^\alpha_\mu$ thought of as the matrix with components $e^\alpha_\mu$.

Before proceeding with the calculation of the anomaly, we  will manipulate $S_E$ such that the coupling to torsion becomes explicit. To achieve this, we break the connection into its Christoffel (torsion-free) and torsionfull parts. Doing so, leads to a fermion-torsion coupling of the form (see App.\ref{App:FTInteractions})
$$
{\cal L}_{int} = {i\over 8}\psi^\dagger K_{\rho\mu\nu}\gamma^\mu[\gamma^\rho,\gamma^\nu] \psi~. 
$$
with $K_{\rho\mu\nu}$ the contorsion tensor defined as\footnote{Note, the upper index in $K$ and $T$ is always lowered to the first position.}
\begin{equation}
    \label{Eq:KTor}
    K_{\rho\mu\nu} = {1\over 2} \left(T_{\rho\mu\nu} + T_{\mu\rho\nu} +  T_{\nu\rho\mu} 
    \right)~~,~~K_{\nu\mu\rho}=-K_{\rho\mu\nu}~.
\end{equation}
We can further simplify ${\cal L}_{int}$ by decomposing $K_{\rho\mu\nu}$ into its irreducible components under rotations. We leave the details for App.\ref{App:FTInteractions} and here we quote the final result 
\begin{equation}
    \label{Eq:TIntAntisym}
    {\cal L}_{int} =  {i \over 64}\psi^\dagger \slashed{S}\gamma_* \psi~,
\end{equation}
where  $S^{\lambda_1\dots\lambda_{n-3}} = {1\over 3!}\epsilon^{\mu\nu\rho\lambda_1\dots\lambda_{n-3}}T_{[\mu\nu\rho]}$ is the Hodge dual of the completely anti-symmetric component of torsion and $\slashed{S}$ its contraction with the completely anti-symmetric product of $n-3$ Dirac matrices (see App. \ref{App:FTInteractions}). Thus, after restricting to $n=4$, we obtain the Dirac operator  with the torsion coupling explicit
\begin{equation}
    \slashed{D}=i \gamma^\mu \left( \partial_\mu + \mathring{\omega}_\mu +iA_\mu + B_\mu \gamma_*\right)~,
    \label{Dir1}
\end{equation}
where $\mathring{\omega}_\mu$ is the Christoffel connection and $B_\mu \equiv A_{5\mu} +\xi S_\mu$ the chiral field capturing the coupling of $\psi$ to torsion and a background U(1) chiral gauge field. We have left the coupling $\xi$ to $S_\mu$ general. The interaction (\ref{Eq:TIntAntisym}) corresponds to $\xi=1/96$.  Note that the divergence of $S_\mu$ is nothing more than the NY invariant.  Therefore, the chiral torsional anomaly is supported entirely by the completely anti-symmetric part of the torsion tensor.

\section{Index}\label{sec:ind}

In this section, we present general remarks regarding the properties of the Dirac operator index and apply them to the case of the chiral torsional anomaly calculation. 

In general, the Dirac operator $\slashed{D}$ is a first order differential operator defined on sufficiently smooth sections $\Gamma(\mathcal{S})$ of a spin bundle $\mathcal{S}$ which satisfy some prescribed boundary conditions at any existing boundaries or singularities. The chirality Dirac matrix $\gamma_*$ acts fiberwise on $\mathcal{S}$ such that $\mathcal{S}=\mathcal{S}_+ \oplus \mathcal{S}_-$ and  has eigenvalues $\pm 1$ on $\mathcal{S}_\pm$, respectively. To be able define an index of $\slashed{D}$, we also request that $\slashed{D}$ anticommutes with $\gamma_*$ and is self-adjoint. If there are boundary conditions, they should commute with $\gamma_*$. Then,
\begin{equation}
    \slashed{D}=\begin{pmatrix} 0 & \mathcal{D}^\dag \\ \mathcal{D} & 0 \end{pmatrix},
\end{equation}
where $\mathcal{D}$ maps positive chirality spinors to the negative chirality ones and $\mathcal{D}^\dag$ is its adjoint. Then one can define an index of $\slashed{D}$ as the difference on numbers of positive and negative chirality zero modes,
\begin{equation}
\label{Eq:Ind1}
    \mathrm{Ind} \slashed{D}=n_+ - n_-.
\end{equation}
This index coincides with the usual analytic index of $\mathcal{D}$, which is the difference between the dimensions of the kernels of $\mathcal{D}$ and $\mathcal{D}^\dag$; ${\rm Ind}\slashed{D}=\dim\, \mathrm{Ker}\, \mathcal{D}-\dim\, \mathrm{Ker}\, \mathcal{D}^\dag$. Let us consider
\begin{equation}
    \slashed{D}^2=\begin{pmatrix}
        \mathcal{D}^\dag \mathcal{D} & 0 \\ 0 & \mathcal{D}\mathcal{D}^\dag 
    \end{pmatrix} .
\end{equation}
Usual arguments show, that nonzero eigenvalues of $\mathcal{D}^\dag \mathcal{D}$ and $\mathcal{D}\mathcal{D}^\dag$ coincide while
$\mathrm{Ker}\, \mathcal{D}=\mathrm{Ker}\, \mathcal{D}^\dag \mathcal{D}$ and $\mathrm{Ker}\, \mathcal{D}^\dag=\mathrm{Ker}\, \mathcal{D} \mathcal{D}^\dag$, see e.g. \cite{Fursaev:2011zz}. Thus we can express the index in terms of $\slashed{D}^2$ as
\begin{equation}
    \mathrm{Ind} \slashed{D} = \mathrm{Tr}\, \left( \gamma_* \exp (-\slashed{D}^2/M^2) \right)~, \label{indM2}
\end{equation}
since the contributions of opposite chiralities non-zero modes cancel against each other, while the zero modes result presicely to ${\rm Ind}\slashed{D}$, Eq. \eqref{Eq:Ind1}. Notably, \eqref{Eq:Ind1} shows that the righthand side of (\ref{indM2}) is independent of $M^2$.

The expression (\ref{indM2}) can be most conveniently analyzed with heat kernel methods. Let $L$ be a Laplace type operator and $Q$ be a smooth matrix valued function. Then there exists a heat kernel (also called a heat trace in the mathematics literature)
\begin{equation}
    K(Q,L,t)=\mathrm{Tr}\left( Q e^{-L/M^2}\right)~, \label{heatk}
\end{equation}
with $M^2\in \mathbb {R}_+$. There is an asymptotic expansion as $M^2\to +\infty$
\begin{equation}
    K(Q,L,M^2)\simeq \sum_{k=0}^\infty M^{n-k}a_k(Q,L), \label{hkexp}
\end{equation}
$n=\dim\MM$.
The coefficients in this expansion, $a_k(Q,L)$, are called the heat kernel coefficients, see \cite{GilkeyNew,Vassilevich:2003xt} for more details.

In this work, we use the expansion (\ref{hkexp}) for $L=\slashed{D}^2$ and $Q=\gamma_* (\delta\phi)$, where $\delta\phi$ is the U(1) chiral transformation parameter. The coefficients $a_k$ then take the form $a_k = \int (\delta\phi)(\dots)$, where the ellipses denote some chiral densities. As was demonstrated in \cite{Obukhov:1997pz} for the case of closed Riemann--Cartan manifolds, the density corresponding to the NY invariant appears in $a_2$. With this in mind, we call the torsional part of the density in $a_2(\gamma_*\delta\phi,\slashed{D}^2)$ a generalized NY density. It will in general depart from the typical NY density, whenever ${\cal M}$ is not a Riemann-Cartan spacetime with empty $\partial {\cal M}$. Therefore, the torsional contribution to the chiral torsional anomaly always appears with a factor of $M^{n-2}$ in the heat kernel expansion. Since ${\rm Ind}\slashed{D}$ is $M$-independent, we arrive at the main conclusion of this Section: \textit{If $\slashed{D}$ admits an index which can be expressed through (\ref{indM2}), the corresponding Nieh--Yan invariant vanishes.}

This conclusion seems to be at odds with the results of \cite{Obukhov:1997pz,Chandia:1997hu} where several examples of manifolds with a nonvanishing NY invariant were constructed. However, a more close inspection of these examples reveals no contradiction. The manifolds in question were described by a line element $\dd s^2=h(r)^2\dd r^2 + f(r)^2\dd\Omega^2$ with $\dd\Omega^2$ being the line element on a 3-sphere and $h$, $f$ arbitrary functions satisfying some mild regularity assumptions. The NY invariant is essentially defined by the value of $f(r)$ at $r\to\infty$, in analogy to magnetic monopoles whose charge is defined by the asymptotic value of a gauge potential. Much depends on the function $h(r)$. If $h(r)$ vanishes sufficiently fast at infinity, the surface $r=\infty$ in fact corresponds to a finite boundary. In this case, one needs to introduce boundary conditions compatible with chirality and take into account boundary contributions to the heat kernel coefficients. We consider this case in detail in the next Section. In the opposite situation, when $h(r)$ is not integrable, we are dealing with a noncompact manifold. In this case, one has to distinguish between local and global heat kernels. If $Q$ has compact support, (\ref{heatk}) and (\ref{hkexp}) exist independently of the global structure of $\MM$, see \cite{berline2003heat}, so that the \emph{local} computations by \cite{Obukhov:1997pz} are correct. If $Q$ does not vanish at infinity as in the case of Index computations with $Q=\gamma_*$, we are dealing with a \emph{global} heat kernel expansion. Even if an infinite volume limit of a particular coefficient $a_k(\gamma_*,\slashed{D}^2)$ (say $a_2(\gamma_*,\slashed{D}^2)$ which gives the NY invariant) does exist, this does not mean that it is related to the index.\footnote{Note, that the authors of \cite{Obukhov:1997pz} never claimed that the NY invariant gives the index.} As we noticed below Eq. (\ref{indM2}), the necessary requirement for the derivation of the heat kernel expression for the index is cancellation between contributions from opposite chiralities. This implies that at least the coefficients $a_2(\tfrac 12(1\pm \gamma_*),\slashed{D}^2)$ should be well-defined in the infinite volume limit, i.e. the chiral torsional anomaly should exist for a single Weyl fermion. This is not the case, however, as may be easily checked by using the expression for torsion from \cite{Obukhov:1997pz} and the heat kernel coefficients in Appendix \ref{sec:hk}. For example, the $B^2$ term gives a divergent contribution to $a_2(\tfrac 12(1\pm \gamma_*),\slashed{D}^2)$. We conclude that the index theorem (\ref{indM2}) is not valid for the geometry considered in \cite{Obukhov:1997pz}. Therefore, while the NY invariant is non-zero, it does not give rise neither to an index nor a chiral torsional anomaly. This argument also shows why we cannot arrive at a finite NY term via Chandia and Zanelli's proposed solution \cite{Chandia:1997hu,Chandia:1997jf,Chandia:1998nu}. To obtain a finite NY term in the index, they proposed an $M$-dependent rescaling of the geometry making $\slashed{D}$ itself $M$-dependent. Since the heat kernel expression does not reproduce the index on any of the rescaled geometries, the rescaling does not save the situation.

A similar comment applies to the singular background torsion considered  in \cite{Rasulian:2023gtb} which led to a delta-function contribution to $\slashed{D}^2$. The authors of \cite{Rasulian:2023gtb} expanded the heat kernel of $\slashed{D}^2$ in powers of this singular contribution and obtained the index depending on the cutoff parameter. However, it is known \cite{Bordag:1999ed} that such expansion does not exists if the delta function has support of codimension greater than 1 and thus cannot be used to evaluate the index. 

\section{Boundaries and chiral torsional anomaly}\label{sec:chian}

In this section, we consider the calculation of the chiral torsional anomaly in spacetimes with boundaries. We fix the boundary conditions supplementing the Dirac operator \eqref{Dir1} and evaluate the chiral torsional anomaly in $n=4$ Euclidean dimensions.

It is known that there are no local boundary conditions which respect chirality and lead to an elliptic boundary value problem \cite{Atiyah:1975jf}. To overcome this difficulty, it was suggested in \cite{Atiyah:1975jf} to consider nonlocal spectral boundary conditions. However, our knowledge of the heat kernel expansion for these conditions is quite limited, see \cite{Gilkey:2004bm}, so that one cannot compute the index or chiral torsional anomaly for spectral boundary conditions. Therefore, we will restrict ourselves to local boundary conditions and to calculations of the chiral torsional anomaly. Since there is no index in this case, there are also no restrictions on the NY term in chiral anomaly.

To be able to compute the chiral torsional anomaly as a response of the effective action to chiral transformation of background fields, we need a Dirac operator which is covariant under chiral transformations supplemented by covariant boundary conditions. The operator (\ref{Dir1}) satisfies this requirement  since it transforms as
\begin{equation}
    \slashed{D}\to \slashed{D}_\phi= e^{\gamma_*\phi}\slashed{D}e^{\gamma_*\phi} \label{chitrD}
\end{equation}
under $B_\mu\to B_\mu + \partial_\mu\phi$. 

A set of local chiral boundary conditions can be constructed in the following way. Let $\MM$ have a smooth boundary $\partial\MM$ and let $\nn$ be an inward pointing unit normal to the boundary. The chiral bag boundary conditions \cite{PhysRevD.12.2733,PhysRevLett.51.1518,Rho:1983bh}, see also \cite{Hrasko:1983sj,Wipf:1994dy}, are defined as
\begin{equation}
\Pi_-(\theta)\psi\vert_{\partial\MM}=0, \qquad \Pi_-(\theta)=\tfrac 12 \bigl( 1- i \gamma_* e^{\gamma_*\theta}\gamma^{\nn}\bigr).\label{Pith}
\end{equation}
Here $\gamma^\nn = n_\mu \gamma^\mu$, $\theta$ is a chiral angle which may depend on the boundary point. One can consider more general boundary conditions by introducing a sign factor in front of $\gamma_*$inside round brackets in (\ref{Pith}), see \cite{Ivanov:2021yms}. It can be checked that for these boundary conditions the normal component of the fermion current $\psi^\dag \gamma^\nn \psi$ vanishes on $\partial\MM$ thus making the Dirac operator selfadjoint. Besides, the conditions (\ref{Pith}) form an orbit of the group of local chiral transformations: The Dirac action is invariant under the transformations (\ref{chitrD}) accompanied by $\psi\to\psi_\phi = e^{-\gamma_*\phi}\psi$ and one can easily check that $\Pi_-(\theta)\psi=0$ implies $\Pi_-(\theta-2\phi)\psi_\phi=0$. Thus, under chiral transformations the angle $\theta$ changes as
\begin{equation}
    \theta\to \theta - 2\phi. \label{chitrtheta}
\end{equation}

Having fixed our boundary conditions, we can now evaluate the chiral torsional anomaly. We do this by employing two distinct regularization schemes, zeta-function and cutoff regularization. The use of these two schemes will exemplify the reasons past literature has thought of the NY term as a regularization artifact.

We begin with the $\zeta$-function regularization approach. The $\zeta$ function of the Dirac operator $\slashed{D}$ with eigenvalues $\lambda$ is defined by the formula
\begin{equation}
\zeta(s,\slashed{D})=\sum_{\lambda>0}\lambda^{-s} + e^{-i \pi s}\sum_{\lambda<0}(-\lambda)^{-s}, \label{zetas}
\end{equation}
where $s$ is a complex parameter. The sums in (\ref{zetas}) are convergent for sufficiently large values of $\Re s$ and the $\zeta$ function can be extended to the whole complex plane as a meromorphic function. The $\zeta$ regularized effective action, whose variation gives us the chiral torsional anomaly, is 
\begin{equation}
W_s=-[\ln\det  {(\slashed{D})}]_s\equiv \Gamma(s)\zeta(s,\slashed{D}).\label{zetaregW}
\end{equation}
The physical point is $s=0$, and $W_0$ requires renormalization to be evaluated. However, the divergent terms are chiral invariant. Thus, chiral variation of the $\zeta$ function regularized action is finite and given by 
\begin{equation}
    \delta_\phi W_{\zeta}=-2 a_4((\delta\phi)\gamma_*,\slashed{D}^2),\label{delphizeta}
\end{equation}
see Appendix \ref{sec:zeta} for details. 

As we can see the $a_2$ heat kernel coefficient and, hence, the NY term is absent from the final result. This would not be the case had we used a cutoff regularization. With the details relegated in App.\ref{sec:zeta}, we can use an integral representation of the $\zeta$ function to define such a regularization. In this case the chiral variation of the regularized action reads
\begin{equation}
    \delta _\phi W_{\Lambda}=-2\sum_{k=0}^4 \Lambda^{4-k} a_k((\delta\phi)\gamma_*,\slashed{D}^2) +\mathcal{O}(\Lambda^{-1})\label{delphiW'}~,
\end{equation}
where $\Lambda$ is the cutoff parameter.

To proceed, we must evaluate the heat kernel coefficients appearing in \eqref{delphizeta}, \eqref{delphiW'}. The heat kernel calculations for arbitrary values of the ``boundary angle'' $\theta$ are quite involved, see \cite{Ivanov:2021yms}. To simplify calculations, we restrict ourselves to $\theta=0$. This allows us to evaluate the chiral torsional anomaly, but forbids us from making two consecutive (infinitesimal) chiral transformations and using the WZ consistency conditions. With this restriction in place, it is straightforward, although tedious, to evaluate the chiral torsional anomaly by using the expressions for the heat kernel coefficients found in Appendix \ref{sec:hk} (see also \cite{Marachevsky:2003zbf}). Below we quote the result in the cutoff regularization \eqref{delphiW'}, since its cutoff-independent part is nothing more than \eqref{delphizeta}. We find
\begin{align}
  \delta _\phi W_{\Lambda}=&-\frac{\Lambda^2}{2\pi^2}\int_{\MM} \dd^4x \sqrt{g} (\delta\phi)\mathring{\nabla}_\mu B^\mu 
  +\frac{\Lambda}{8\pi^{3/2}}\int_{\partial\MM} \dd^3x \sqrt{h}(\delta\phi)\mathring{\nabla}_\mu B^\mu 
  \nonumber
  \\
  &+\int_{\MM} \dd^4x \sqrt{g} (\delta\phi)\Biggl[-{i \over 48\pi^2} \epsilon^{\mu\nu\rho\sigma}\left( B_{\mu\nu}B_{\rho\sigma} -{1\over 2}\mathring{R}_{\mu\nu\lambda\theta}\mathring{R}\indices{_{\rho\sigma}^{\lambda\theta}}\right) 
  \nonumber
  \\
  &+ {1\over 12\pi^2}\left(\mathring{G}^{\mu\nu}+ g^{\mu\nu}\mathring{\nabla}^2\right)\mathring{\nabla}_\mu B_\nu -{2\over 3\pi^2}\left(B^\mu B^\nu+{1\over 2}g^{\mu\nu}B^2\right)\mathring{\nabla}_\mu B_\nu\Biggr]
  \nonumber
  \\
  &- \int_{\partial\MM} \dd^3x \sqrt{h}\left[\delta\phi\left({i\over 30\pi^2}\epsilon^{abc}B_b\mathring{\nabla}_a B_c + {2\over 15\pi^2}B^2 B_\nn\right)+{1\over 12\pi^2}(\nabla_\nn \delta\phi) \mathring{\nabla}_\mu B^\mu\right]
  \label{delphiW}~,
\end{align}
where $B_{\mu\nu} = \mathring{\nabla}_{[\mu}B_{\nu]}$, $\mathring{G}^{\mu\nu} = R^{\mu\nu} -g^{\mu\nu}\mathring{R}/2$ is the Einstein tensor, the letter $a$, $b$, $c$ are used to denote vector indices tangential to the boundary, $\sqrt{h}$ is the induced volume element on $\partial\MM$, $\epsilon^{abc} = n_\mu \epsilon^{\mu abc}$, and we have ignored mixed chiral-vector anomaly terms. The bulk terms in the expression above are consistent with earlier calculations \cite{Obukhov:1982da,Cognola:1987wd,Yajima:1987eh} while the cutoff independent boundary terms reconfirm the calculations in \cite{Marachevsky:2003zbf} with nonabelian axial vector fields. For interested readers, we like to mention a recent extensive study of boundary anomalies \cite{FarajiAstaneh:2023fad} which however did not include torsion.

There are several interesting features in Eq.~\eqref{delphiW}. First, we observe that apart from the bulk NY term contribution, we also find a boundary one, which diverges as ${\cal O}(\Lambda)$. As expected from our result in the zeta function regularization scheme, these divergent terms are an artifact and can be cancelled via suitable counterterms. More precisely, we cancel them by adding to the effective action the following local gauge invariant counterterms
\begin{equation}
    W_{\mathrm{c.t.}}=-\frac{\Lambda^2}{4\pi^2}\left[ \int_{\MM} \dd^4x \sqrt{g} B^2 -\int_{\partial\MM}\dd^3x \sqrt{h} \theta B_\nn  \right] + \frac{\Lambda}{16\pi^{3/2}}\int_{\partial\MM} \dd^3x \sqrt{h}\theta\mathring{\nabla}_\mu B^\mu~.
\end{equation}
We remind that (\ref{delphiW}) was computed for $\theta=0$. Thus one has to put $\theta=0$ after computing the chiral variation of counterterms. The fact we can cancel the NY term exactly, implies that there is an interaction we can add to the Dirac action, which upon renormalization of the coupling constant negates the NY term. In this sense, both the bulk and boundary NY terms are irrelevant \cite{Bilal:2008qx}. 

Second, the bulk terms in the second line of \eqref{delphiW} are the Pontryagin densities associated to the chiral gravitational anomaly and the chiral anomaly of $B_\mu$, respectively. In terms of the decomposition $B_\mu = A^5_\mu + \xi S_\mu$ discussed below equation \eqref{Dir1}, we can write
\begin{align}
    \label{Eq:PntrTorsion}
    \epsilon^{\mu\nu\rho\sigma}B_{\mu\nu}B_{\rho\sigma}  &= \epsilon^{\mu\nu\rho\sigma}\left(A_{\mu\nu}A_{\rho\sigma} + 2\xi A_{\mu\nu} S_{\rho\sigma} +\xi^2 S_{\mu\nu}S_{\rho\sigma}\right)
    \nonumber 
    \\
    &= \epsilon^{\mu\nu\rho\sigma}\left(A_{\mu\nu}A_{\rho\sigma} + 2\xi A_{\mu\nu} \mathring{\nabla}_\kappa T^\kappa_{\rho\sigma} +\xi^2\mathring{\nabla}_\lambda  T^\lambda_{\mu\nu}\mathring{\nabla}_\kappa T^\kappa_{\rho\sigma}\right)~, 
\end{align}
where $A_{\mu\nu} = \mathring{\nabla}_{[\mu}A^5_{\nu]}$ and similarly for $S_{\mu\nu}$. In the last equality, we have restored the torsion tensor, as this expression makes clear some features of the Pontryagin term. In particular, since we are taking derivatives with respect to the torsionless Christoffel connection, the highest power of the torsion tensor appearing in \eqref{Eq:PntrTorsion} is two and no term can be directly related to the NY invariant.
Therefore, the Pontryagin density of $B_\mu$ resolves into the Pontryagin density for $A^5_\mu$, a mixed anomaly between $A^5_\mu$  and the natural divergence of the torsion tensor, $\mathring{\nabla}_\mu T^\mu_{\nu\rho}$, and the Pontryagin density for the Hodge dual of torsion. To the extend that $A^5_\mu$ and torsion can be delineated experimentally, since they are sourced by different fields, these three anomalies can also be isolated and studied independently. While we know that the $A_{\mu\nu}$ Pontryagin term leads to the chiral magnetic effect \cite{fukushima2008chiral}, we are not aware of any studies on the effects of the other two terms to anomalous electron transport. While some no-go results exist for anomalous torsional transport \cite{Ferreiros:2020uda, Chernodub:2021nff}, the Pontryagin term \eqref{Eq:PntrTorsion} evades their assumptions by being higher-order in derivatives. Thus, it would be interesting to perform the analysis found in \cite{Ferreiros:2020uda} by expanding their sets of assumptions to include \eqref{Eq:PntrTorsion}. Assumptions similar or identical to \cite{Ferreiros:2020uda} are made in the cond-mat analysis of the torsional anomaly, as far as we can tell, which explains why they have missed the torsional Pontryagin term. This is not true for the high-energy literature, where several authors have reported this term (see e.g. \cite{Kreimer:1999yp}).

We move on to the term on the third line of \eqref{delphiW}. This term might look like a bulk term, but we can express it as a boundary term up to a total chiral variation. To be precise, we have
\begin{align}
    &\int_{\MM} \dd^4x \sqrt{g} (\delta\phi)\Biggl[{1\over 12\pi^2}\left( \mathring{G}^{\mu\nu}+ g^{\mu\nu}\mathring{\nabla}^2\right)\mathring{\nabla}_\mu B_\nu -{2\over 3\pi^2}\left(B^\mu B^\nu+{1\over 2}g^{\mu\nu}B^2\right)\mathring{\nabla}_\mu B_\nu\Biggr]
    \nonumber
    \\
    &=\delta_\phi\left[{-1\over 24\pi^2}\int_{\MM} \dd^4x \sqrt{g}B^\mu\left(\mathring{G}_{\mu\nu}\mathring{R} +\mathring{\nabla}_\mu\mathring{\nabla}_\nu-2B^2g_{\mu\nu}\right)B^\nu\right]
    \nonumber 
    \\
    &+\int_{\partial\MM} \dd^3x \sqrt{h}\left[\delta\phi{1\over 12\pi^2}\left(\mathring{G}_{\nn\nu} +\mathring{\nabla}_\nn\mathring{\nabla}_\nu-4B^2\nn_\nu\right)B^\nu + {1\over 24\pi^2}\left(B_\nn \mathring{\nabla}^2\delta\phi-\mathring{\nabla}_\mu B^\mu\mathring{\nabla}_\nn \delta\phi\right)\right]~.
    \label{Eq:BulktoBoundary}
\end{align}
We can absorb the chiral variation term into the definition of the effective action $W_\Lambda$, which amounts to a choice of renormalization scheme. As an important corollary, note that in spacetimes without boundaries, only the Pontryagin term \eqref{Eq:PntrTorsion} contributes to the torsional anomaly.  This makes our results consistent with \cite{Ferreiros:2020uda}, where no anomalous torsional transport was found for zeroth order derivative torsion contributions to the anomaly. Thus, we see that the inclusion of boundaries is not only essential to obtain a non-trivial NY invariant, but also to obtain non-trivial chiral transport, when torsion is homogeneous. 

Let us then focus on the boundary contributions to the anomaly. These read, after combining \eqref{delphiW} and \eqref{Eq:BulktoBoundary},
\begin{align}
    \delta_\phi W_{\Lambda,\partial\MM}={1\over 60\pi^2}\int_{\partial\MM} \dd^3x \sqrt{h}\Bigl[&\delta\phi\left( -{5\over 2} g_{\nn\mu} \mathring{R}_3 + 5\mathring{\nabla}_\nn \mathring{\nabla}_\mu - 80\overset{\leftarrow}{\mathring{\nabla}}_\nn \mathring{\nabla}_\mu +  (30\overset{\leftarrow~~}{\mathring{\nabla}^2}-336 B^2)n_\mu \right)B^\mu
    \nonumber 
    \\
    &+{i\over 3}\delta\phi\epsilon^{abc}B_{[a}\mathring{\nabla}_b B_{c]} \Bigr]~,
    \label{Eq:BoundaryAnom}
\end{align}
where the left arrow over the covariant derivatives means they act to the left. We have also used that for vanishing extrinsic curvature, $G_{\nn\mu} = -g_{\nn\mu} \mathring{R}_3/2$, with $\mathring{R}_3$ the boundary curvature \cite{Gourgoulhon:2007ue}.

\noindent The last term in \eqref{Eq:BoundaryAnom} is a Chern-Simons term for $B_\mu$. This term is a chiral variation of 
\begin{equation}
    W_{\rm CS} =-{i\over 360\pi^2}\int_{\partial\MM} \dd^3x \sqrt{h}  \theta \epsilon^{abc}B_{[a}\mathring{\nabla}_b B_{c]}~.
    \label{Eq:CSBoundary}
\end{equation}
The $W_{\rm CS}$ contribution to the anomaly, implies the existence of a torsional anomalous Hall effect on $\partial\MM$: Consider two copies of the present system with boundary angles $\theta_0$ and $\theta_0 + \delta\theta$, respectively, and glue them at their boundary.\footnote{Our results are valid up to ${\cal O}(\theta^2)$ terms, thus in what follows we assume the constant angles $\theta_0$ and $\delta\theta$ are $o(\theta^2)$.} This setup reproduces the one found in topological insulators \cite{Sekine:2020ixs}, where $B_\mu$'s role is played by an electromagnetic gauge field. In contrast though to a topological insulator, a Chern-Simons term is induced by the gauge variation of a \textit{bulk} axion term;\footnote{Schematically, the topological insulator Lagrangian reads $\phi P(F)$, where $\phi$ is the axion and $P(F)$ the Pontryagin density of the electromagnetic field.} we only need the $\theta$ field to be well-defined on the boundary. This avoidance of the bulk also seems to allow evading a quantization condition on the corresponding Hall conductivity. The boundary angles $\theta_0$ or $\theta_0 +\delta\theta$ do not have to be a multiple of $\pi$ and the corresponding anomalous torsional Hall conductivity is not quantized. In particular, the boundary current induced by the discontinuity in $\theta$ reads
\begin{equation}
    \left.j^a_5\right|_{\partial \MM} = {\delta\theta \over 180\pi^2}i\epsilon^{abc}\mathring{\nabla}_{[b}B_{c]}\equiv \sigma_{T}i\epsilon^{abc}\mathring{\nabla}_{[b}B_{c]}~,
    \label{Eq:BoundaryCurrent}
\end{equation}
where we have identified $\sigma_T =\delta\theta/180\pi^2$ with the anomalous torsional Hall conductivity.\footnote{The factor of $i$ was not included in $\sigma_T$, as it will be absorbed by $\epsilon^{abc}$ upon Wick rotating back to spacetime.} Equation \eqref{Eq:BoundaryCurrent} and the expression for $\sigma_T$ is one of our main results for anomalous torsional transport. 

A direct relationship between the remaining boundary terms (first line of \eqref{Eq:BoundaryAnom}) and their effect on transport currently eludes us, but what we can say for certain is that they lead to non-trivial anomalous contributions to the $2-$ and $4-$point functions of the chiral current. They also affect the $3-$point function between two chiral currents and the energy-momentum tensor. Interestingly, these higher-point functions are between a single bulk operator and operators confined on the boundary. This suggests that bulk spin and chiral charge can be ``transmuted'' to boundary spin and momentum, leading to variants of the anomalous torsional Hall effects found in $(2+1)$-dimensions \cite{Parrikar:2014usa}, and discussed above. It would be interesting to explore these effects further in future work. 

Finally, regarding the $T^2$ temperature dependence of the NY term reported in the literature \cite{Nissinen:2019mkw,Nissinen:2019wmh,Huang:2019haq, Chernodub:2021nff}. We have not considered temperature-dependent contributions to the chiral torsional anomaly, but where they to appear in the final result, they would appear along side the $\Lambda^2$ contribution. This does not imply that the temperature contribution can be absorbed via a suitable redefinition of the counterterms, however. This is because the temperature, being the length of the thermal circle, is a non-local quantity and, hence cannot appear in the expression for local counterterms. Therefore, the thermal NY contribution seems to be a bona fide anomaly.

\section{Conclusions}\label{sec:concl}

We have brought to the fore, once more, the debate in the literature regarding the chiral torsional anomaly in general and the presence and well-posedeness of the NY contribution in particular. We have isolated the confusion to the presence or not of a relationship between the index of the torsionful Dirac operator, the anomaly, and the Nieh-Yan invariant. 

We re-examined the derivation of the heat kernel formula (\ref{indM2}) for the index of a Dirac operator. Our first observation is that the mathematical conditions for having a well-defined index differ from the conditions for having a well-defined anomaly. For the index to exist, the operator $\slashed{D}$ should anticommute with $\gamma_*$, while the domain of $\slashed{D}$ should be invariant under the action of $\gamma_*$. This latter requirement implies that boundary conditions commute with $\gamma_*$ and excludes all local boundary value problems for a single Dirac fermion. To use the heat kernel formula (\ref{indM2}) for the index, one has to make sure that the heat kernel expansion (\ref{hkexp}) exists and that the heat kernel coefficients are integrable over the whole manifold. This excludes singularities inside $\MM$ and requires sufficiently fast falloff of background fields at infinity.  We conclude thus that the index is ill-defined in the examples considered in \cite{Chandia:1997hu,Rasulian:2023gtb}.

We find that the NY term contribution to both the bulk and boundary can always be cancelled by a local counterterm. This shows that the NY term is not a genuine anomaly, as it  can be absorbed into the renormalization of the torsion couplings. This does not mean that the NY invariant is irrelevant to Dirac fermion physics, but that its effects on transport should be accompanied by a careful discussion of renormalization conditions.

Not all is lost, however, since apart from the NY term several additional contributions to the chiral torsional anomaly are present. In the bulk, we found a Pontryagin density contribution for the Hodge dual of torsion, or equivalently for the divergence, $\mathring{\nabla}_\mu T^\mu_{\nu\rho}$, of torsion (see Eq.~\eqref{delphiW}). This term had eluded previous condensed matter calculations in the literature, as it is higher order in derivatives than that usually considered. Our bulk term is not new from the high-energy point of view.

What is new however is the proof that most of the bulk terms can be re-expressed as boundary ones, in addition to novel boundary terms. There is a mixture of topological and non-topological terms in the boundary. The non-topological terms depend explicitly on the metric and its curvature, and up to second derivatives of torsion and the chiral variation parameter. Their interpretation in terms of transport is not clear, but they are expected to allow for the interplay between bulk and boundary transport of chiral charge, spin and momentum. The topological term is a Chern-Simons term for the Hodge dual of the torsion tensor, which we show can be written as the chiral variation of a  boundary action $W_{\rm CS}$ (see Eq. \eqref{Eq:CSBoundary}). This term implies the existence of the torsional anomalous Hall effect for samples with a discontinuous boundary angle. We presented an expression for the torsional Hall conductivity as a function of this discontinuity in Eq.~\eqref{Eq:BoundaryCurrent}.

We may, of course, accept the point of view that the cutoff $\Lambda$ is fundamental in defining the UV completion of the theory and thus no renormalization is needed. However, in such a case one still finds additional terms in the chiral torsional anomaly besides the NY term, both in the bulk and on the boundary. These terms are in principle subleading to the NY term, since $\Lambda$ is larger than other scales associated with background bosonic fields. However, since there is nothing special about the NY term, a consistent approach should treat all terms on equal footing.

Besides settling part of the debate regarding the chiral torsional anomaly, our work also opens several avenues for future research. It would be interesting to understand how the non-topological boundary terms we found affect torsional transport. This can be achieved by evaluating the Ward identities they lead to between the bulk chiral current and boundary current or energy-momentum operators.  It would also be interesting to understand the redefinition ambiguity between the chiral torsional and chiral vortical effects noted in \cite{Ferreiros:2020uda} on the microscopic level, in particular how to independently tune torsion and Riemannian curvature on the level of such effective models of electron dynamics in solids. This ambiguity is based on the possibility to absorb torsion into Riemannian curvature by a redefinition of the Christoffel connection, without affecting the total spacetime curvature; schematically $R = \mathring{R}+ \mathring{\nabla}T+ T^2$ (see e.g. \cite{Erdmenger:2022nhz, Erdmenger:2023hne}). It would be interesting to understand how this manifests itself microscopically in e.g. tight binding models. Finally, it would be interesting to clarify the role of the chiral angle $\theta$. From the technical point of view, the introduction of this angle was necessary to make the boundary conditions chiral covariant and allows to represent the anomaly as a chiral variation of a functional determinant. It is not clear yet whether $\theta$ is a dynamical field and whether it can be identified as a collective fermionic degree of freedom. In this regard, our torsional Chern-Simons term $W_{CS}$ \eqref{Eq:CSBoundary} anomalous contribution might be of help. In analogy to the chiral anomaly, where a jump in the $\theta$ angle leads to an anomaly inflow onto the domain wall which is cancelled by the parity anomaly of Dirac fermions on the domain wall, $W_{CS}$ on the domain wall seems to imply a similar picture in connection to the chiral torsional anomaly. We will explore these and further aspects of torsion in the solid state context in future works. 

\begin{center}
\section*{Acknowledgments} 
\end{center}

We thank Bastian He{\ss} for collaboration at an early stage of this project. 
IM's work is supported by the STFC consolidated grant (ST/X000583/1) ``New Frontiers In Particle Physics, Cosmology And Gravity''.
DV was supported in part by the S\~ao Paulo Research Foundation (FAPESP), project 2021/10128-0, and by the National Council for Scientific and Technological Development (CNPq), project 304758/2022-1. JE and RM acknowledge funding by Germany's Excellence Strategy through the W\"urzburg‐Dresden Cluster of Excellence on Complexity and Topology in Quantum Matter ‐ ct.qmat (EXC 2147, project‐id 390858490),
and  by the Deutsche Forschungsgemeinschaft (DFG) 
through the Collaborative Research centre ``ToCoTronics'', Project-ID 258499086—SFB 1170.

\appendix
\medskip

\section*{Appendix}
\numberwithin{equation}{section}

\section{Heat kernel coefficients}\label{sec:hk}
In the literature, the heat kernel coefficients are commonly written for generic operators of Laplace type
\begin{equation}
    L=-(\nabla^2+E) \label{genL}
\end{equation}
where $\nabla_\mu =\partial_\mu + \bar{\omega}_\mu$ is a covariant derivative while $E$ is a bundle endomorphism (a matrix valued function). The operator $\slashed{D}^2$ takes the form (\ref{genL}) with 
\begin{eqnarray}
    &&\bar{\omega}_\mu = \mathring{\omega}_\mu +iA_\mu +\tfrac 12 [\gamma_\mu,\gamma_\nu] B^\nu \gamma_* ~,\label{omegamu}
    \\
    &&E=-\tfrac 14 \mathring{R} +\tfrac i4 [\gamma_\mu,\gamma_\nu] F^{\mu\nu}+\mathring{\nabla}_\mu B^\mu \gamma_* +2B_\mu B^\mu \label{E}~.
    \end{eqnarray}
We will also need   
\begin{equation}
    \Omega_{\mu\nu} \equiv [\nabla_\mu,\nabla_\nu ]= \gamma^\kappa \mathring{\nabla}_{[\nu}B_{|\kappa|}\gamma_{\mu]}\gamma_*-B_{\mu\nu}\gamma_* +{i\over 4}\gamma^\kappa\gamma^\tau \mathring{R}_{\kappa\tau\mu\nu}+ i \gamma^\kappa \gamma_{[\nu}\gamma^\tau\gamma_{\mu]}B_\kappa B\tau~.
    \label{AppEq:Omega}
\end{equation}
The vertical bars denote that the enclosed index is omitted from anti-symmetrization and $B_{\mu\nu} = \mathring{\nabla}_{[\mu}B_{\nu]}$.

\noindent With a simplifying assumption of vanishing extrinsic curvature of $\partial\MM$, one has \cite{Marachevsky:2003zbf,Kurkov:2018pjw}
\begin{eqnarray}
&&a_0(Q,L)=\frac 1{(4\pi)^{n/2}} \int_{\mathcal{M}} d^nx \sqrt{g}\, \mathrm{tr}\, Q\,,\label{a0}\\
&&a_1(Q,L)=\frac 1{4(4\pi)^{(n-1)/2}} \int_{\partial\MM} d^{n-1}x \sqrt{h}\, \mathrm{tr}\, (Q\chi)\,,\label{a1}\\
&&a_2(Q,L)=\frac 1{6(4\pi)^{n/2}} \left[ \int_{\mathcal{M}} d^nx \sqrt{g}\, \mathrm{tr}\, Q
\left(6 E +  R\right) 
 + \int_{\partial\MM} d^{n-1}x \sqrt{h}\, \mathrm{tr}\, 3\nabla_\nn Q  \right] ,\label{a2} \\
&&a_3(Q,L)=\frac 1{384(4\pi)^{(n-1)/2}} \int_{\partial\MM} d^{n-1}x \sqrt{h}\, \mathrm{tr}\, \left[
Q\bigl( -24E +24\chi E \chi +48\chi E +48 E\chi \right. \nonumber\\
&&\qquad \qquad  \left. +16\chi R - 8\chi R_{j\nn}^{\ \ j\nn} -12\nabla_j \chi \nabla^j \chi +12\nabla_j\nabla^j \chi)
 +24\chi \nabla_\nn \nabla_\nn Q
\right] \label{a3}
\\
&&a_4(Q,L) = {1 \over 360 (4\pi)^{n/2}}\Biggl\lbrace \int_{\mathcal{M}} d^nx \sqrt{g}\, \mathrm{tr}\, \biggl[Q\bigl(60 \nabla^2 E + 60 RE + 180E^2 + 30\Omega_{\mu\nu}^2 
\nonumber
\\
&& \qquad \qquad + 12\nabla^2 R + 5 R^2-2R_{\mu\nu}^2 + 2R_{\mu\nu\rho\sigma}^2\bigr)  \biggr]  
\nonumber
\\
&&\qquad \qquad +\int_{\partial\MM} d^{n-1}x \sqrt{h}\, \mathrm{tr}\, \biggl[Q\bigl(30\nabla_\nn E + 30\chi\nabla_n E \chi + 90\chi \nabla_\nn E + 90 \nabla_\nn E\chi
\nonumber 
\\
&&\qquad\qquad +18\chi\chi_{:a}\Omega_{an}+ 12
\chi_{:a}\Omega_{an}\chi +
  18 \Omega_{an}\chi\chi_{:a}-  12 \chi\Omega_{an}\chi_{:a}  + 6 [\chi\Omega_{an}\chi, \chi_{:a}] 
  \nonumber 
  \\
  &&\qquad\qquad+ 54
[\chi_{:a}, \Omega_{an}] + 30[\chi, \nabla^a \Omega_{a\nn}] + 12 \nabla_\nn R + 30 \chi_{:aa}
 - 6 \chi_{:a}\chi_{:a}\chi + 30\chi \nabla_\nn R\bigr) 
 \nonumber 
 \\
 &&\qquad\qquad+ \nabla_\nn Q\bigl(-30E + 30 \chi E\chi + 90 \chi E  +90 E\chi -18 \chi_{ :a} \chi_{ :a}   +30 \chi R\bigr)+30\chi \nabla^2\nabla_\nn Q\biggr] \Biggr\rbrace .
\end{eqnarray}

Here $\chi=i\gamma_*\gamma_\nn$ is the matrix which defines boundary projector through $\Pi_-=\tfrac{1}{2}(1-\chi)$ for $\theta=0$ and $\chi_{:a} = \nabla_a\chi$. Note that the replacement $\chi\to i \gamma_* e^{\theta\gamma_*}\gamma_\nn$ does not yield correct heat kernel coefficients for $\theta\neq 0$, see \cite{Esposito:2005dn}.

\section{Zeta function regularization and anomaly}\label{sec:zeta}

In the present appendix, we discuss the zeta function regularization for the heat kernel and derive the expressions \eqref{delphizeta} and \eqref{delphiW'} for the chiral anomaly. 

Let us separate the parts of regularized effective action (\ref{zetaregW}) which are even/odd with respect to the overall sign inversion of $\slashed{D}$
\begin{equation}
    W_s=W_{s,\mathrm{even}}+W_{s,\mathrm{odd}} \label{evenodd}
\end{equation}
where $W_{s,\mathrm{even/odd}}=\tfrac 12( W_s(\slashed{D})\pm W_{s}(-\slashed{D}))$. At $s=0$, $W_{s,\mathrm{odd}}$ gives the parity anomaly which is not considered in this work. So, in what follows we only consider $W_{s,{\rm even}}$ and drop the subscript even.

 After some algebra one obtains
\begin{equation}
    W_{s}=\tfrac 12 \Gamma(s) \bigl(1+e^{-i \pi s}\bigr) \zeta(s/2,\slashed{D}^2)\label{Weven}~.
\end{equation}

The heat kernel and the $\zeta$ function are related through a Mellin transform,
\begin{equation}
    \zeta(s,L,Q)=\frac{1}{\Gamma(s)} \int_0^\infty \dd t\, t^{s-1} K(Q,L,t), \label{zetaK}
\end{equation}
where the smeared $\zeta$ function is defined as
$
     \zeta(s,L,Q)=\mathrm{Tr}\left( Q L^{-s}\right)
$.
The singularity structure and some particular values of the $\zeta$ function are defined by the heat kernel coefficients.
\begin{equation}
    \mathop{\mathrm{Res}}_{s=\frac{n-k}{2}}\bigl( \Gamma(s)\zeta(s,L,Q)\bigr) = a_k (Q,L) \label{Reszeta}
\end{equation}
In particular,
$
    \zeta(0,L,Q)=a_n(Q,L) 
$.

Let us define a double regularization of the effective action by introducing a cutoff in the integral representation (\ref{zetaK}) of the $\zeta$ function.
\begin{equation}
    \zeta_\Lambda (s/2,\slashed{D}^2) =\frac{1}{\Gamma(s/2)}\int_{\Lambda^{-2}}^\infty \dd t\, t^{\frac s2 -1} \mathrm{Tr}\left( e^{-t\slashed{D}^2}\right)~.\label{zetaLambda}
\end{equation}
To get back the $\zeta$ regularized effective action one should take $\Lambda\to\infty$ before an analytic continuation to $s=0$. In the cutoff regularization, the order of operations is opposite.

\noindent The variation of (\ref{zetaLambda}) under infinitesimal chiral transformation $\delta_\phi\slashed{D}=(\delta\phi)\gamma_*\slashed{D}+\slashed{D}(\delta\phi)\gamma_*$ reads
\begin{equation}
    \delta_\phi \zeta_\Lambda (s/2,\slashed{D}^2) =\frac{4}{\Gamma(s/2)}\int_{\Lambda^{-2}}^\infty \dd t\, t^{\frac s2}\partial_t \mathrm{Tr}\left( (\delta\phi)\gamma_* e^{-t\slashed{D}^2}\right)~.
\end{equation}
We can re-express the integral in a useful form by integrating by parts to obtain
\begin{equation}
   \delta_\phi \zeta_\Lambda (s/2,\slashed{D}^2) =-\frac{4}{\Gamma(s/2)} \left[ \int_{\Lambda^{-2}}^\infty \dd t\, t^{\frac s2 -1}
   \left( \frac{s}{2}\right)\mathrm{Tr}\left( (\delta\phi)\gamma_* e^{-t\slashed{D}^2}\right)
    +\Lambda^{-s} \mathrm{Tr}\left( (\delta\phi)\gamma_* e^{-\slashed{D}^2/\Lambda^2}\right) \right]\label{delzeta2'}
\end{equation}
There is no contribution from the upper limit since the heat kernel decays exponentially fast at $t\to\infty$. We substitute (\ref{delzeta2'}) in (\ref{Weven}) and continue to $s=0$ to obtain
\begin{equation}
    \delta_\phi W_{\Lambda}=-2\, \mathrm{Tr}\left( (\delta\phi)\gamma_* e^{-\slashed{D}^2/\Lambda^2}\right)
\end{equation}
With the help of the heat-kernel expansion (\ref{hkexp}) as a power series in $\Lambda$, this expression can be rewritten as an expansion in $\Lambda$, which gives Eq.(\ref{delphiW'}) quoted in the main text.

In contrast, in the $\zeta$ function regularization we immediately put $\Lambda^{-1}=0$ and then integrate by parts
\begin{equation}
   \delta_\phi\zeta (s/2,\slashed{D}^2) =-\frac{4}{\Gamma(s/2)} \left[ \int_{0}^\infty \dd t\, t^{\frac s2 -1}
   \left( \frac{s}{2}\right)\mathrm{Tr}\left( (\delta\phi)\gamma_* e^{-t\slashed{D}^2}\right)
    +t^{\frac{s}{2}} \mathrm{Tr}\left( (\delta\phi)\gamma_* e^{-\slashed{D}^2/\Lambda^2}\right)\vert_{t=0} ~.\right]\label{delzeta2}
\end{equation}
The integral gives $\zeta(s/2,(\delta\phi)\gamma_*,\slashed{D}^2)$ while the last term vanishes since the calculation must be done for large positive $\Re s$. Thus, one arrives at Eq.\ (\ref{delphizeta}).

\section{Fermion-torsion interaction in $n$ dimensions}\label{App:FTInteractions}

In this appendix, we make the coupling of fermions to torsion explicit. To achieve this, we assume that the spin connection is not an independent field, but is given in terms of the vielbeine and the ``vector" connection $\Gamma^{\lambda}_{\mu\nu}$ via
\begin{equation}
    \label{AppEq:LinToSpinCon}
    \omega^{\alpha}_{\beta\mu} = -E^\nu_\beta\partial_\mu e^\alpha_\nu + E^\nu_\beta\Gamma^{\lambda}_{\mu\nu}e^\alpha_\lambda~.
\end{equation}
We decompose $\Gamma^{\lambda}_{\mu\nu}$ into its Christoffel, $\mathring{\Gamma}^{\lambda}_{\mu\nu}$, and non-Christoffel, $K^\lambda_{\mu\nu}$ parts. This decomposition allows us to split $\omega^\alpha_{\beta\mu}$ as 
\begin{equation}
    \omega^{\alpha}_{\beta\mu} = (-E^\nu_\beta\partial_\mu e^\alpha_\nu + E^\nu_\beta\mathring{\Gamma}^{\lambda}_{\mu\nu}e^\alpha_\lambda) + E^\nu_\beta K^{\lambda}_{\mu\nu}e^\alpha_\lambda\equiv \mathring{\omega}^\alpha_{\beta\mu} + E^\nu_\beta K^{\lambda}_{\mu\nu}e^\alpha_\lambda ~.
\end{equation}
The connection $\mathring{\omega}^\alpha_{\beta\mu}$ is the spin connection in the absence of torsion, while $K^{\lambda}_{\mu\nu}$ describes all torsion contributions. Namely,
\begin{equation}
    \label{AppEq:TorK}
     T^\lambda_{~~\mu\nu} = K^\lambda_{~~\mu\nu
    } - K^\lambda_{~~\nu\mu}~.
\end{equation}
We can invert \eqref{AppEq:TorK} in a metric spacetime, whence $K_{\rho\mu\nu} = - K_{\nu\mu\rho}$. In particular, 
\begin{equation}
    \label{AppEq:KTor}
    K_{\rho\mu\nu} = {1\over 2} \left(T_{\rho\mu\nu} + T_{\mu\rho\nu} +  T_{\nu\rho\mu} 
    \right)~.
\end{equation}
With the above definitions in place for the connection, the coupling of a Dirac fermion to torsion is given by the term
$$
{\cal L}_{int} = {i\over 8}\psi^\dagger K_{\rho\mu\nu}\gamma^\mu[\gamma^\rho,\gamma^\nu] \psi~. 
$$

We could stop our discussion here, but the Clifford algebra satisfied by the Dirac matrices allows for additional simplifications. Namely, in $n$ dimensions, we can decompose the product of three Dirac matrices into irreducible representations of $SO(n)$. To do so, we use the identity
\begin{equation}
    \label{AppEq:AntiCom}
    \gamma^\mu [\gamma^\rho, \gamma^\nu] = {1 \over 2}\lbrace \gamma^\mu, [\gamma^\rho, \gamma^\nu] \rbrace + 2\delta^{\mu[\rho}\gamma^{\nu]}~.
\end{equation}
The first term in \eqref{AppEq:AntiCom} is completely anti-symmetric under interchange of its indices, while the second is clearly independent of the first. Thus, \eqref{AppEq:AntiCom} splits the Dirac matrix product in irreducible representations of $SO(n)$. Therefore, because of the orthonormality of irreps, contracting \eqref{AppEq:AntiCom} with the contorsion tensor will give us the decomposition we are after. In particular, the torsion interaction reads
\begin{align}
    {\cal L}_{int} &= {i \over 64}\psi^\dagger K_{[\rho\mu\nu]}\lbrace \gamma^\mu, [\gamma^\rho, \gamma^\nu] \rbrace \psi+ {i \over 2}\psi^\dagger \gamma^\nu K\indices{^\rho_\rho_\nu}\psi 
    \nonumber
    \\
    &= {i \over 128}\psi^\dagger T_{[\rho\mu\nu]}\lbrace \gamma^\mu, [\gamma^\rho, \gamma^\nu] \rbrace \psi+ {i \over 2}\psi^\dagger \gamma^\nu T\indices{^\rho_\rho_\nu}\psi~,
\end{align}
where in the last equality we used \eqref{AppEq:KTor}. Let us focus on the completely antisymmetric part of torsion. We choose to write it in terms of its Hodge dual, which we define as 
\begin{equation}
    T_{[\mu\nu\rho]} = {1\over (n-3)!}\epsilon_{\mu\nu\rho\lambda_1\lambda_2\dots\lambda_{n-3}}S^{\lambda_1\lambda_2\dots\lambda_{n-3}} \Leftrightarrow S^{\lambda_1\dots\lambda_{n-3}} = {1\over 3!}\epsilon^{\mu\nu\rho\lambda_1\dots\lambda_{n-3}}T_{[\mu\nu\rho]}~.
\end{equation}
To make use of the Hodge dual of torsion, we also need the Hodge dual of the completely antisymmetric combination of the Dirac matrices. Up to an overall sign, this is given by \cite{Freedman_VanProeyen_2012}
\begin{equation}
\epsilon_{\mu\nu\rho\lambda_1\lambda_2\dots\lambda_{n-3}} \lbrace \gamma^\mu, [\gamma^\rho, \gamma^\nu] \rbrace = 2(n-3)! \gamma_{\lambda_1\dots\lambda_{n-3}}\gamma_*~,
\end{equation}
where $\gamma_{\lambda_1\dots\lambda_{n-3}}$ is the completely anti-symmetrized product of $n-3$ Dirac matrices \cite{Freedman_VanProeyen_2012}. 

\noindent Finally, we have for the torsion interaction in terms of the dual field
\begin{equation}
    \label{AppEq:GenDInteraction}
    {\cal L}_{int} =  {i  \over 64}\psi^\dagger \slashed{S}\gamma_* \psi+ {i \over 2}\psi^\dagger \gamma^\nu T\indices{^\rho_\rho_\nu}\psi~,
\end{equation}
with $\slashed{S}$ the obvious contraction between the $S_{\lambda_1\dots}$ field and $\gamma^{\lambda_1\dots}$.

\noindent The second term in \eqref{AppEq:GenDInteraction}, proportional to the torsion trace would seem to break unitarity. It can, however, be shown that this term can be removed by using an explicitly real Dirac Lagrangian (see e.g. \cite{Ferreiros:2020uda}). Therefore, the final form of the torsion-fermion interaction Lagrangian is
\begin{equation}
    \label{AppEq:TIntAntisym}
    {\cal L}_{int} =  {i \over 64}\psi^\dagger \slashed{S}\gamma_* \psi~.
\end{equation}
For $n=4$, we find the result quoted in the main text.

\bibliographystyle{utphys}
\bibliography{refs}

\end{document}